# Epitaxial aluminum layer on antimonide heterostructures for exploring Josephson junction effects

W. Pan[1], K.R. Sapkota[2], P. Lu[2], A.J. Muhowski[2], W.M. Martinez[2], C.L.H. Sovinec[2], R. Reyna[2], J.P. Mendez[2], D. Mamaluy[2], S.D. Hawkins[2], J.F. Klem[2], L.S.L. Smith[1,3], D.A. Temple[3], Z. Enderson[4,5], Z. Jiang[5], and E. Rossi[6]

[1]Sandia National Labs, Livermore, California 94550, USA
[2]Sandia National Labs, Albuquerque, New Mexico 87123, USA
[3]Department of Physics, Norfolk State University, Norfolk, Virginia 23504, USA
[4]Oak Ridge Institute for Science and Education Postdoctoral Fellowship, Oak Ridge, Tennessee 37831, USA
[5]School of Physics, Georgia Institute of Technology, Atlanta, Georgia 30332, USA
[6]Department of Physics, William & Mary, Williamsburg, Virginia 23187, USA

Abstract

In this article, we present results of our recent work of epitaxially-grown aluminum (epi-Al) on antimonide heterostructures, where the epi-Al thin film is grown at either room temperature or below zero $^{o}$C. A sharp superconducting transition at T ~ 1.3 K is observed in these epi-Al films. We further show that supercurrent states are realized in Josephson junctions fabricated in the epi-Al/antimonide heterostructures with mobility $\mu \sim 1.0\times10^{6}$ $cm^2/Vs$. These results clearly demonstrate we have achieved growing high-quality epi-Al/antimonide heterostructures, a promising platform for the exploration of Josephson junction effects for quantum information science and microelectronics applications.

Aluminum (Al) as a superconductor is widely used in Josephson junctions (JJ) [1,2] and superconducting high Q resonators [3]. In recent years, much work has been devoted to epitaxially-grown Al (epi-Al) on semiconductors for achieving an interface of the highest quality between Al and the semiconductor, which should help realize topological superconductivity [4,5] and Majorana zero modes [6-8]. Examples include epi-Al on InAs or InSb nanowires for hard superconducting gaps [9,10], epi-Al on InAs quantum wells (QWs) [11-14] for planar JJs [15], and epi-Al on Si for quantum computing applications [16].

So far, less work has been carried out on epi-Al on antimonide heterostructures, in particular InAs/GaSb type-II heterostructures. InAs/GaSb is a unique material system [17]. Depending on the thickness of InAs and GaSb quantum wells, the material can be a conventional semiconductor, a zero-band gap semimetal, or an inverted-band gap quantum spin Hall insulator [18,19]. The latter is of particular interest, since Majorana zero modes can be achieved at zero magnetic field in JJs mediated by the edge channels in a quantum spin Hall insulator [20]. This will simplify the design of topological qubits [21] and offer great promise of eventually achieving fault-resistant universal quantum computing.

Herein we first present results of our recent work of epitaxially-grown Al on InAs/GaSb type-II heterostructures and InAs QWs. A schematic experimental flow of structure and transport characterization for the measurements described in the main text is shown in Fig. S1 in Supplemental Materials (SM) [22]. In both samples, a sharp superconducting transition at T ~ 1.3K is observed. We then present results for a JJ fabricated in the second epi-Al on InAs QW heterostructures, where the starting growth temperature of the epi-Al layer is below 0 °C. For this epi-Al heterostructure, an InAs QW structure that can achieve electron mobility $\mu \sim 1.0\times10^6$ $cm^2/Vs$ is used. Supercurrents states are observed in this JJ. Moreover, near the critical current, a multi-switching behavior is observed. An effective two-junctions model is proposed to explain this behavior. We want to notice here that the focus of this manuscript is on the growth and characterization of epi-Al layer. The purpose of including the preliminary results of a Josephson junction is to show that the epi-Al can induce supercurrent states in a Josephson junction based on our epi-Al antimonide heterostructures. Detailed studies of Josephson effects in optimized planar Josephson junctions are undergoing and results will be reported in future publications. Nevertheless, the observed large mobility, combined with the strong spin-orbit coupling of InAs, and the evidence of a superconducting proximity effect, make the structure an ideal candidate for the realization of non-Abelian states useful for quantum computations, given that disorder (low mobility) has emerged as the main obstacle toward the robust establishment and clear detection of such states.

In this work, two epi-Al heterostructures are grown. The heterostructure in Fig. 2a is based on the previous studies published in Refs. [18] and [19], in which the quantum spin Hall insulator phase is expected to exist. The heterostructure in Fig. 2b is based on the report of achieving the highest electron mobility [23]. In growing both heterostructures, an unintentionally doped GaSb wafer is used as the substrate for all the materials growth. The growth sequence of first heterostructure (Fig. 2a) is as follows. After a GaSb buffer layer (600 nm thick) and a short-period superlattice of GaSb/AlSb QWs, a 200-nm-thick $Al_{0.33}Ga_{0.67}Sb$ potential barrier layer is grown. For the first structure the growth is then followed by InAs (13 nm) and GaSb (5 nm) double quantum wells and an AlSb (10 nm) potential barrier layer. At this point, effusion cells are brought to idle temperature and the substrate heater is turned off. After cooling the wafer completely overnight to room temperature, the Al cell is reheated. Then, the final 30-nm-thick Al layer is grown on top of the AlSb layer. The growth temperature for each epi-layer is given in Fig. S1 in SM [22]. For the second structure (Fig. 2b), after the GaSb buffer layer, superlattices layer, and $Al_{0.33}Ga_{0.67}Sb$ bottom potential barrier, an InAs QW of 21 nm thick is grown, followed by an AlSb (20nm) potential barrier and a 5 nm thick GaSb cap layer. In growing the epi-Al layer in this second heterostructure, the substrate is rotated towards the cryo-pump so that the substrate temperature is lowered below 0 °C before the Al layer growth [11-14, 24-26]. We point out here that in epi-Al growth a low growth temperature helps avoid migration of Al [11-14, 24-26]. Moreover, a low growth temperature helps minimize the strain effect caused by lattice mismatch between the Al and antimonide semiconductor. We further point out here that in recent years materials growth/synthesis and devices fabrication at low temperatures have become more and more important in achieving new functionalities and architectures, and in preserving the bulk properties without significant degradation. Example includes cryogenic epi-growth or deposition of superconducting materials such as Al [11-14, 24-26] and Ta [27] on semiconductor heterostructures.

The first epi-Al heterostructure is structurally characterized using aberration-corrected scanning transmission electron microscopy (AC-STEM) using an FEI Titan™ G2 80-200 STEM with a Cs probe corrector and ChemiSTEM™ technology, operated at 200keV. TEM samples are prepared using a Thermo Fisher Helios G-5 plasma FIB (PFIB) with $Xe^{+}$ ions at a temperature of -160 °C. The cryogenic PFIB condition is to avoid the FIB induced AlGa epi-growth artifact, which could happen unintentionally [28]. High-angle annular dark-field (HAADF) STEM imaging is used for structural analysis. More details can be found in Ref. [28].

Fig. 1a shows a cross section STEM image. It is evident that the Al layer is continuous but not uniform. Instead, large Al islands are formed. This is consistent with atomic force microscopy (AFM) results (Fig. 1b). Here, AFM scans are carried out in an area of 5×5 $\mu m^2$ to examine the surface morphology. Watershed masking through the software Gwyiddion is performed to isolate and separate the Al islands. The average equivalent disc radius is 54.7 nm. The surface roughness values are calculated by standard surface roughness equations and performed on a larger scan representative of the full dataset, and an RMS roughness of 5.86 nm is deduced. Fig. 1c shows a HAADF STEM image of the heterostructures in Fig. 1a. Perfect epitaxial interfaces are seen between InAs, GaSb, and AlSb, respectively. Fig. 1d shows a zoomed image for the interface between Al and AlSb. It is clearly seen that Al forms epitaxially with the AlSb layer underneath, with orientations Al(110)//AlSb(001) and Al[1,−1,0]//AlSb[1,−1,0]. This epitaxial relationship is determined by analyzing the high-resolution STEM image and accompanying fast-Fourier transform (FFT) pattern in the top-right inset.

A specimen of size ~ 5×5 $mm^2$ is then cleaved from each wafer for electronic transport characterization. Indium is placed along the edges and corners for ohmic contacts. Low frequency (~ 11 Hz) phase lock-in amplifier technique is used for the sample resistance measurements. For the DC current-voltage (I-V) characterization of Josephson junction (Fig. S3 in Ref. [22]), a Keithley 238 source meter is also used. All low temperature measurements are carried out in a pumped $^3$He system.

Figs. 2c and 2d show the four-terminal resistance $R_{xx}$ as a function of temperature for the epi-Al layers in the two heterostructures, respectively. In the first heterostructure where epi-Al is grown at room temperature, the resistance is nearly constant, ~ 1 Ω, at high T. At T ~ 1.3 K, $R_{xx}$ drops sharply and reaches the zero-resistance value, when the Al thin film becomes superconducting. The range of temperature over which $R_{xx}$ drops is only 20 mK, indicating a high quality of the epi-Al film and uniform transition temperature across the whole specimen. The results on the temperature dependence of critical magnetic field are shown in Fig. S4 in SM [22].

In the second heterostructure, where the epi-Al layer is grown below 0 °C, again a superconducting transition is observed around 1.3 K (Fig. 2d). Different from the epi-Al layer grown at room temperature, the normal-state resistance is extremely low, about 0.12 Ω. This lower resistance is likely due to an improved uniformity of the Al film (Fig. S5 in [22]) grown at lower temperatures. $R_{xx}$ as a function of B field is shown in Fig. S6 [22], from which a critical magnetic field $B_c$ ~ 5 mT is obtained.

We further note here that the growth structure of the second heterostructure is based on the one used in Ref. [23] to achieve the highest electron mobility of two-dimensional gas (2DEG) in an InAs QW. To characterize the quality of the 2DEG in our growth, an identical InAs QW structure without epi-Al is grown (Fig. S7a), and the results of low temperature magneto-transport characterization of the 2DEG are shown in Fig. S7b [22]. Well-developed Shubnikov-de Haas oscillations and integer quantum Hall effect are observed in the low and high B regimes, respectively, demonstrating the high quality of the InAs QW. An electron density of $n = 5.4 \times 10^{11}$ $cm^{-2}$ and mobility $\mu \sim 1.0\times10^6$ $cm^2/Vs$ are obtained from the $R_{xx}$ and $R_{xy}$ data. These values are comparable to those reported in InAs two-dimensional gas (2DEG) also grown on GaSb substrates [23, 29-31], corroborating the high quality of our antimonide heterostructures. Using an effective mass of 0.03 $m_e$ [23], we deduce an electron mean free path $L_{mfp}$ = 12 μm and Fermi velocity of $v_F= 7.1\times10^5$ m/s.

The results above clearly show that we have achieved high quality Al epi-growth on InAs/GaSb type-II heterostructures and InAs QWs. In the following, we will present results in JJs made of epi-Al on the second heterostructure (Fig. 2b).

JJs are fabricated in this second epi-Al-antimonide heterostructure. We use photoresist developer MF319 to etch off the Al in the junction region. An SEM image of a JJ is shown in Fig. 3a. The length of the junction (i.e., the separation between the two Al electrodes) L is about 2 μm, and the width 12.5 μm. Using the $v_F$ value obtained above and an Al superconducting gap of $\Delta_{Al} = 1.76k_BT_c \approx 0.23$ meV ($T_c \approx 1.3$K), a BCS superconducting coherence length [32] $\xi_0 \equiv \hbar v_F/(\pi\Delta_{Al}) \approx 0.64$ μm is deduced.

Furthermore, it is clearly seen that the color and the morphology of the junction region are different from that of the two large Al electrodes (Fig. 3b). The darker color in the region is most likely due to the static electromagnetic effect from the electron beam. EDS (at an energy level of 10 kV) analysis (Fig. 3c) corroborates that aluminum is etched off in the junction region. Indeed, the shape of the junction is evident in the Al EDS image, while for In, Ga, and Sb a uniform distribution is displayed. We attribute the Al noise in the background to the underlying Al present in AlSb and AlGaSb layers. Additionally, we measure the depth of the etch to be ~ 40 nm, indicating that both the epi-Al and top GaSb layers are removed. This leaves the 2DEG in the InAs QW as the only conduction channel between the two Al electrodes. Finally, a much larger junction resistance in the normal state (~ 12 Ω, Fig. 3d) compared to that of the epi-Al layer (Fig. 2d) also corroborates the conclusion that aluminum is etched off in the junction region.

The temperature dependence of the junction resistance $R_{xx}$ displays two superconducting transitions (Fig. 3d). The first occurs at $T_c$ ~ 1.5 K when the epi-Al thin film become superconducting. It is slightly higher than the bulk Al thin film superconducting transition temperature of 1.3 K (Fig. 2d). This larger $T_c$ is probably due to the epi-Al electrodes becoming more amorphous [33] through device fabrication, as demonstrated in Fig. 3b. The second transition occurs at ~ 1.2 K, presumably when the 2DEG channel in the junction itself becomes superconducting.

The ability to induce superconducting correlations into an InAs QW with such high mobility is an important advance toward the realization of topological superconducting states supporting non-Abelian states given that disorder has emerged as the main obstacle toward the attainment of such states [34]. Given the potential of non-Abelian states for the realization of fault tolerant qubits and gates, the availability of superconductor-semiconductor heterostructure based on InAs QWs with such a large mobility would be an important contribution toward quantum computing [35] and superconducting microelectronics [36].

The current-voltage (I-V) curve of the JJ is measured at zero magnetic field and shown in Fig. 3e. The current is swept in the up direction. Overall, the I-V curve behaves as expected in a typical JJ. In the large current regime (I > 120 μA), the I-V curve is linear, and the sample is in the normal state. Below a critical current of $I_c$ ~ 105 μA, the dc voltage across the two Al electrodes $V_{dc}$ is zero, and the JJ is in the supercurrent states.

Examining the data carefully, it is clearly seen that, near the critical current when the voltage jumps to a non-zero value, a multi-switching behavior is observed (Fig. 3e and Fig. S8 in [22]). This multi-switching behavior can be understood by invoking a model of an effective two-Josephson-junctions (Junctions A and B, in Fig. 3e and Fig. 4a) with different critical currents and normal-state resistances. For Junction A, the critical current is $I_c^A$ ~ 105 μA and the normal state resistance is $R_N^A$ ~ 0.8 Ω (obtained from a linear fit of the data points marked by the red open squares). For Junction B, the values are $I_c^B$ ~ 5 μA and $R_N^B$ ~ 8 Ω, respectively. As the current is swept from 0, at beginning it flows almost exclusively through Junction A, due to a much larger critical current this junction can accommodate. When the current reaches a value larger than $I_c^A$, Junction A becomes non-superconducting and a sizable portion of the current starts to flow through Junction B, which is still superconducting. This redirection of current may bring the current through Junction A below its critical value, and Junction A becomes superconducting again/ Consequently, a certain amount of the current now flows back to Junction A, until Junction A becomes normal again. This dynamic process of current from Junction A to B and then back A is consistent with the switching between the low resistance state and the high resistance state observed between ~ 105 and ~ 115 μA, see Fig. 3e. Such switching behavior stops after the current is large enough and both Junctions A and B become normal.

Beyond this point, the I-V curve becomes linear and there is no multi-switching behavior, as observed in Fig. 3e. We further notice here that the normal state resistance when both Junctions A and B are non-superconducting is smaller than that (~ 12 Ω) in Fig. 3d, suggesting that in the current regime we have measured the Al electrodes are still superconducting.

The time-averaged voltage-current (V-I) characteristic for the scenario above can be semi-quantitatively described using an effective RSJ (resistively shunted junction) model, as depicted in Fig. 4a, where the cross (X) represents an ideal Josephson junction and the solid rectangular the junction's resistance. For an individual junction, the V-I characteristic can be written as $V = R_N \times I \times \Theta(I - I_c)$. Here $\Theta(x)$ is a step function, $\Theta(x) = 0$ for $x \leq 0$ and $\Theta(x) = 1$ for $x > 0$. The V-I trace for Junction A (B) without Junction B (A) is shown as the dashed line in Fig. 4b. With two junctions co-existing (Fig. 4a), the total voltage $V = R_N^A \times I \times \Theta(I - I_c^A) + R_N^B \times I \times \Theta(I - I_c^A - I_c^B)$, as marked by the solid line. The estimated values of $R_N^A \sim 0.8\ \Omega$, $I_c^A \sim 105\ \mu A$, $R_N^B \sim 8\ \Omega$, and $I_c^B \sim 5\ \mu A$ are given above. The placing of the second junction in series with the resistive channel of the first junction, allows the RSJ model to capture, on average, the effect of the current's dynamical switching between the two superconducting channels when $I_c^A < I < I_c^A + I_c^B$, as described above. The overall shape of the V-I obtained with the proposed RSJ model agrees well with the experimental observations.

Below, we argue how the two effective junctions can form in our device structure. As shown in Fig. 3a, the separation between the two Al electrodes at the edges (~ 2 μm) is quite shorter than that in the central region (~ 6.5 μm). Consequently, we can expect that the region close to the edges will form one junction (Junction A) with a large critical current (105 μA) and a small normal-state resistance (0.8 Ω), while the central part will form a junction (Junction B) with a small critical current (5 μA) and a large normal-state resistance (8 Ω). With these assignments, we calculate the values of $eI_cR_N$ for the two junctions, and find them to be ~ 84 and 40 μeV, respectively. These values are smaller than the Al superconducting gap of $\Delta_{Al}$ ~ 230 μeV, indicating a low interface transparence between the epi-Al layer and InAs quantum well.

In summary, we have achieved high quality epi-growth of aluminum on antimonide heterostructures. A sharp superconducting transition is observed in our epi-Al thin films. We further show that the epi-Al antimonide heterostructures are a promising material platform for exploring Josephson effects for superconducting microelectronics and quantum information applications.

We thank J. Cuozzo for helpful discussions and L.F. Schnebly for taking the SEM images in Figure S5. The work at Sandia is supported by the LDRD program at Sandia National Laboratories. W.P. and E.R. acknowledges support from DOE, Grant No. DE-SC0022245. A portion of this work was performed at the National High Magnetic Field Laboratory, which is supported by National Science Foundation Cooperative Agreement No. DMR-2128556 and the State of Florida. Device fabrication was performed at the Center for Integrated Nanotechnologies, a U.S. DOE, Office of BES, user facility. Sandia National Laboratories is a multi-mission laboratory managed and operated by National Technology & Engineering Solutions of Sandia, LLC (NTESS), a wholly owned subsidiary of Honeywell International Inc., for the U.S. Department of Energy's National Nuclear Security Administration (DOE/NNSA) under contract DE-NA0003525. This written work is authored by an employee of NTESS. The employee, not NTESS, owns the right, title and interest in and to the written work and is responsible for its contents. Any subjective views or opinions that might be expressed in the written work do not necessarily represent the views of the U.S. Government. The publisher acknowledges that the U.S. Government retains a non-exclusive, paid-up, irrevocable, world-wide license to publish or reproduce the published form of this written work or allow others to do so, for U.S. Government purposes. The DOE will provide public access to results of federally sponsored research in accordance with the DOE Public Access Plan.

Figures and figure captions

Figure 1

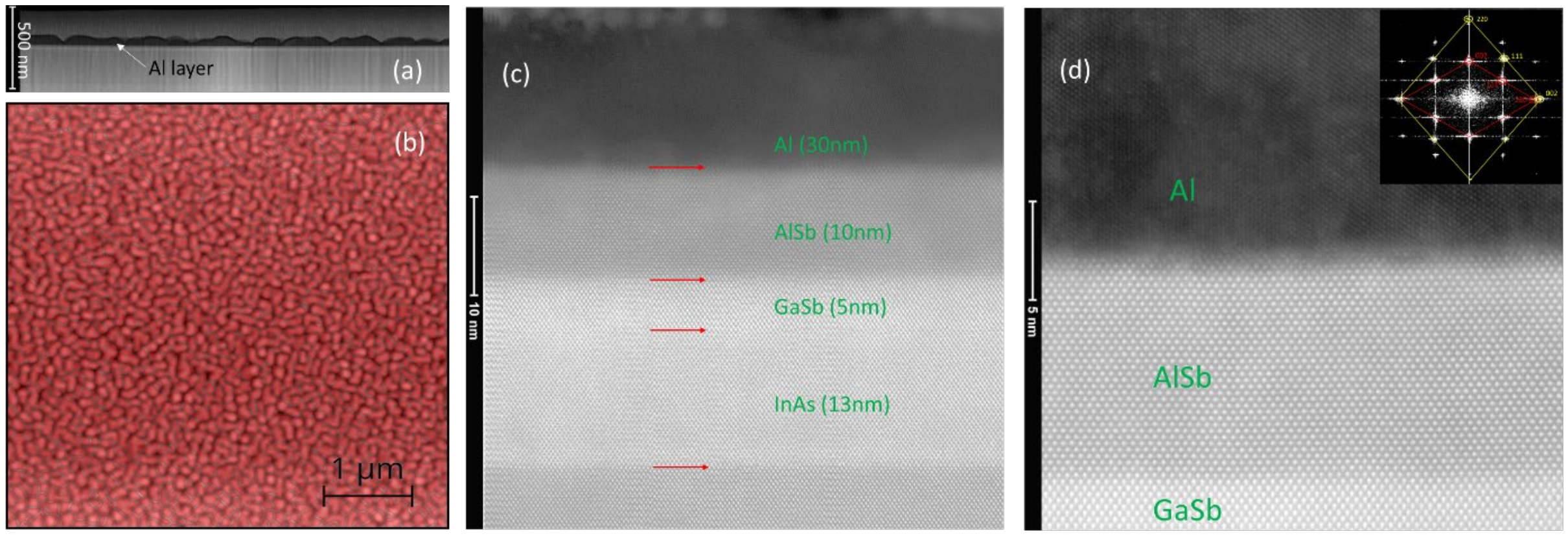

Fig. 1 Structural and electronic characterizations of epi-Al thin film on an AlSb/GaSb/InAs heterostructure. (a) Cross-section image of the epi-Al layer, the scale bar is 500nm. (b) AFM image of epi-Al film over a 5×5 $\mu m^2$ area. (c) HAADF STEM image of the heterostructures in Figure 1a. The scale bar is 10 nm. (d) High-resolution STEM image of Al and AlSb epitaxial relationship. The scale bar is 5 nm. The inset shows the accompanying fast-Fourier transform pattern.

Figure 2

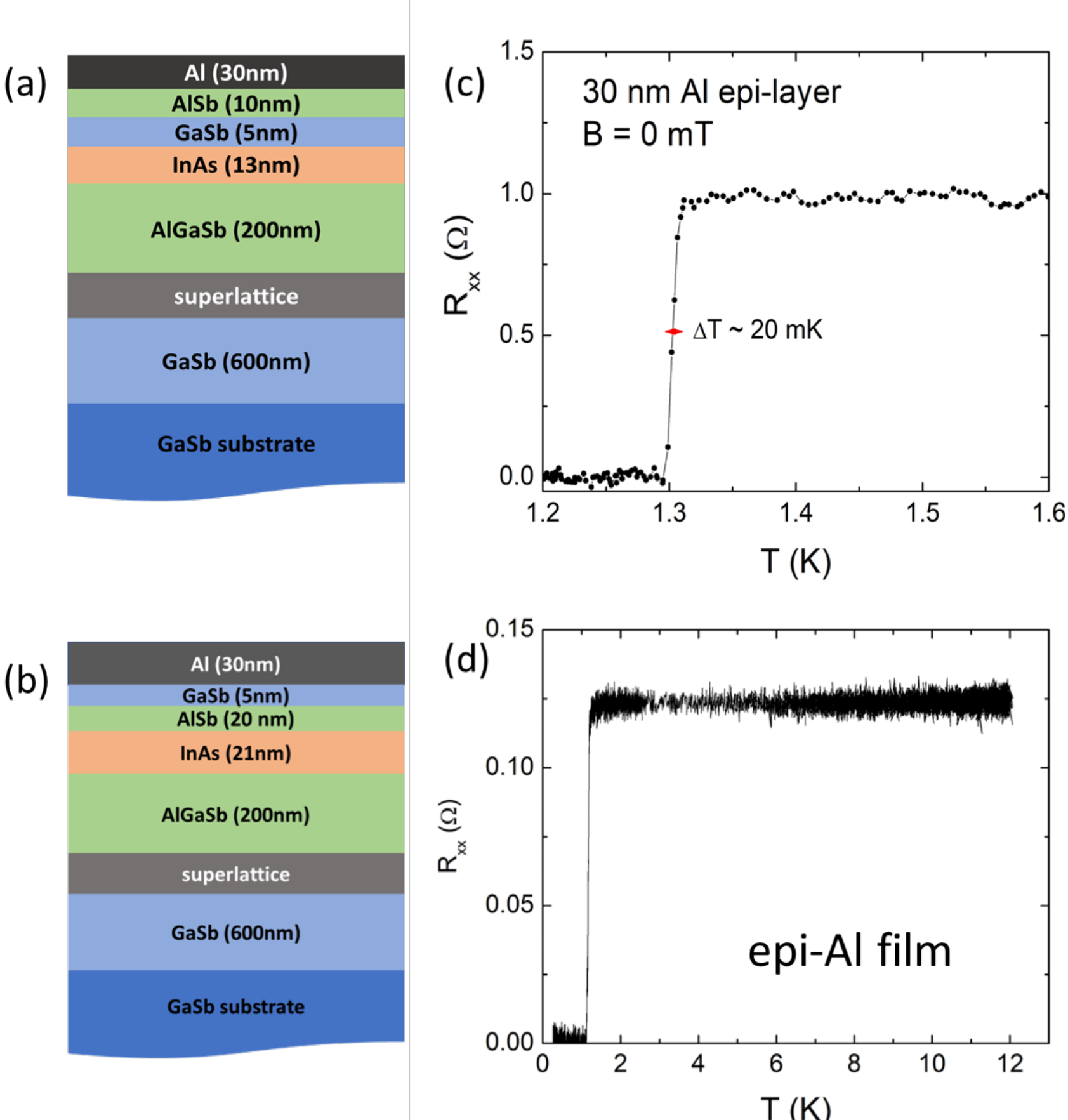

Fig. 2 (a) and (b) Growth structures of two epi-Al heterostructures as discussed in the main text. (c) and (d) Sample resistance $R_{xx}$ as a function of temperature at B = 0 T in the two epi-Al layers, respectively. A superconducting transition is seen at T ~ 1.3 K in both epi-Al layers.

Figure 3

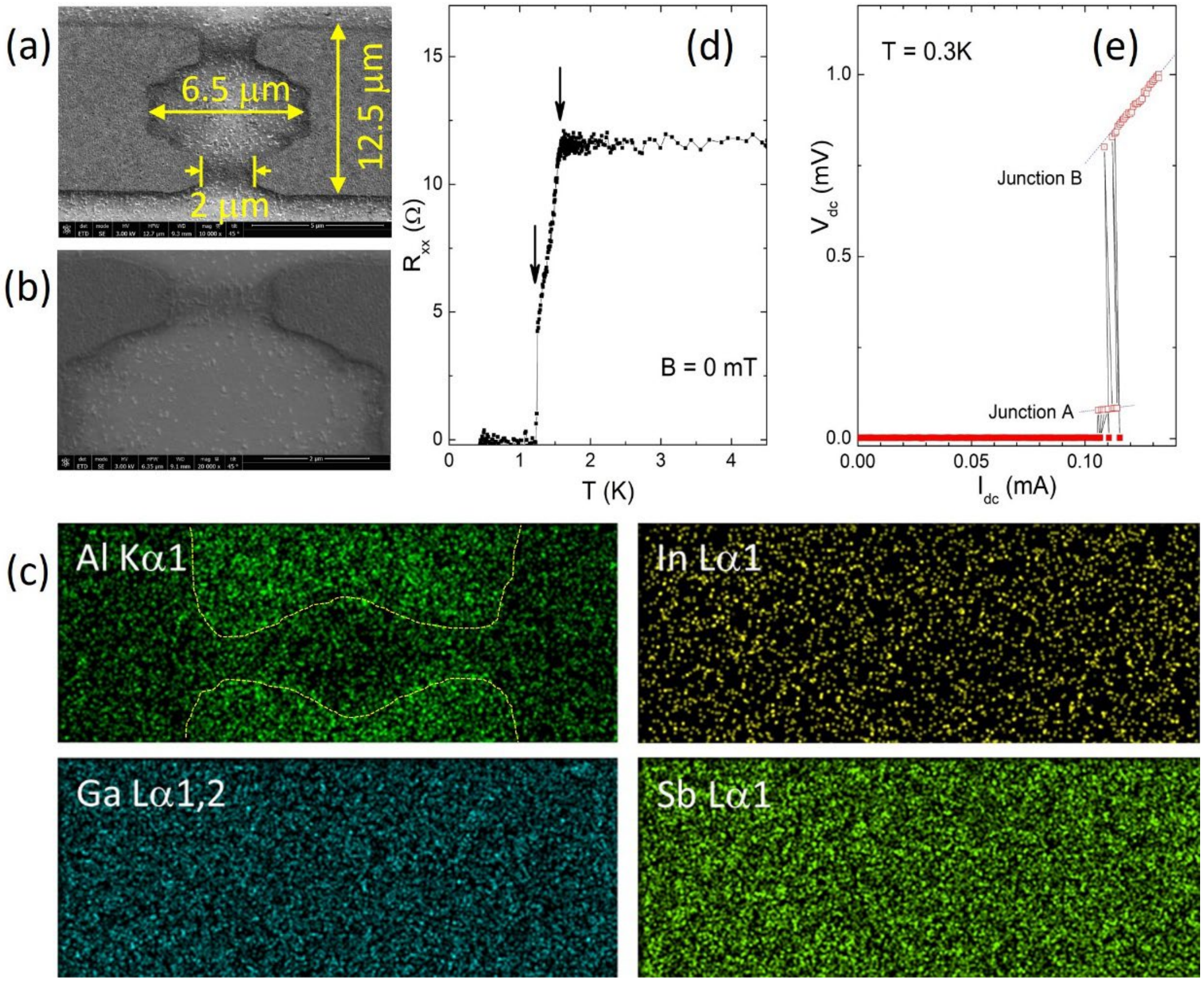

Fig. 3 (a) SEM image of a large Josephson junction (JJ). (b) Zoomed image of Fig. 3(a). (c) EDS maps of Al Kα, Ga Lα, In Lα, and Sb Lα, extracted from spectral image with selected EDS energy windows for each element. The dotted green line is a guide to the eye. (d) Superconducting transition in the JJ, marked by the arrows. (e) I-V curves for current sweeping up from zero voltage. The dashed lines are liner fits, from which the normal resistance is obtained for Junction A and B.

Figure 4

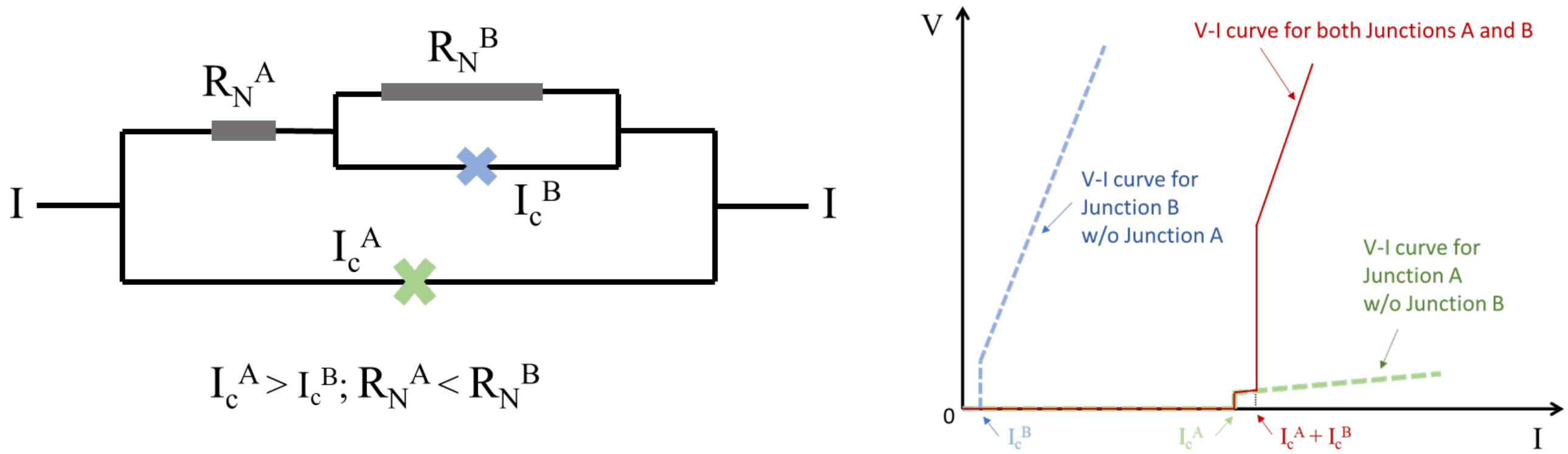

Fig. 4 (a) Effective RSJ model for our device with two Josephson junctions. The cross (X) represents a Josephson junction and the solid rectangular the junction resistance. (b) Schematic of V-I curves. The green/blue dashed line represents the V-I curve for individual Junction A/B, respectively. The total V-I curve for the two co-existing junctions in Fig. 4(a) is depicted by the red solid line.

Supplemental Materials

# Epitaxial aluminum layer on antimonide heterostructures for exploring Josephson junction effects

W. Pan[1], K.R. Sapkota[2], P. Lu[2], A.J. Muhowski[2], W.M. Martinez[2], C.L.H. Sovinec[2], R. Reyna[2], J.P. Mendez[2], D. Mamaluy[2], S.D. Hawkins[2], J.F. Klem[2], L.S.L. Smith[1,3], D.A. Temple[3], Z. Enderson[4,5], Z. Jiang[5], and E. Rossi[6]

[1]Sandia National Labs, Livermore, California 94550, USA
[2]Sandia National Labs, Albuquerque, New Mexico 87123, USA
[3]Department of Physics, Norfolk State University, Norfolk, Virginia 23504, USA
[4]Oak Ridge Institute for Science and Education Postdoctoral Fellowship, Oak Ridge, Tennessee 37831, USA
[5]School of Physics, Georgia Institute of Technology, Atlanta, Georgia 30332, USA
[6]Department of Physics, William & Mary, Williamsburg, Virginia 23187, USA

Our epi-Al antimonide heterostructures are by MBE. in situ RHEED (reflection high energy electron diffraction) is used to monitor the thickness of layers. The base pressure is xxxx before growth. In Fig. S1, we show the starting growth temperature for each layer in the first heterostructure. The growth temperature is the same for the same compound semiconductor layer in the second heterostructure.

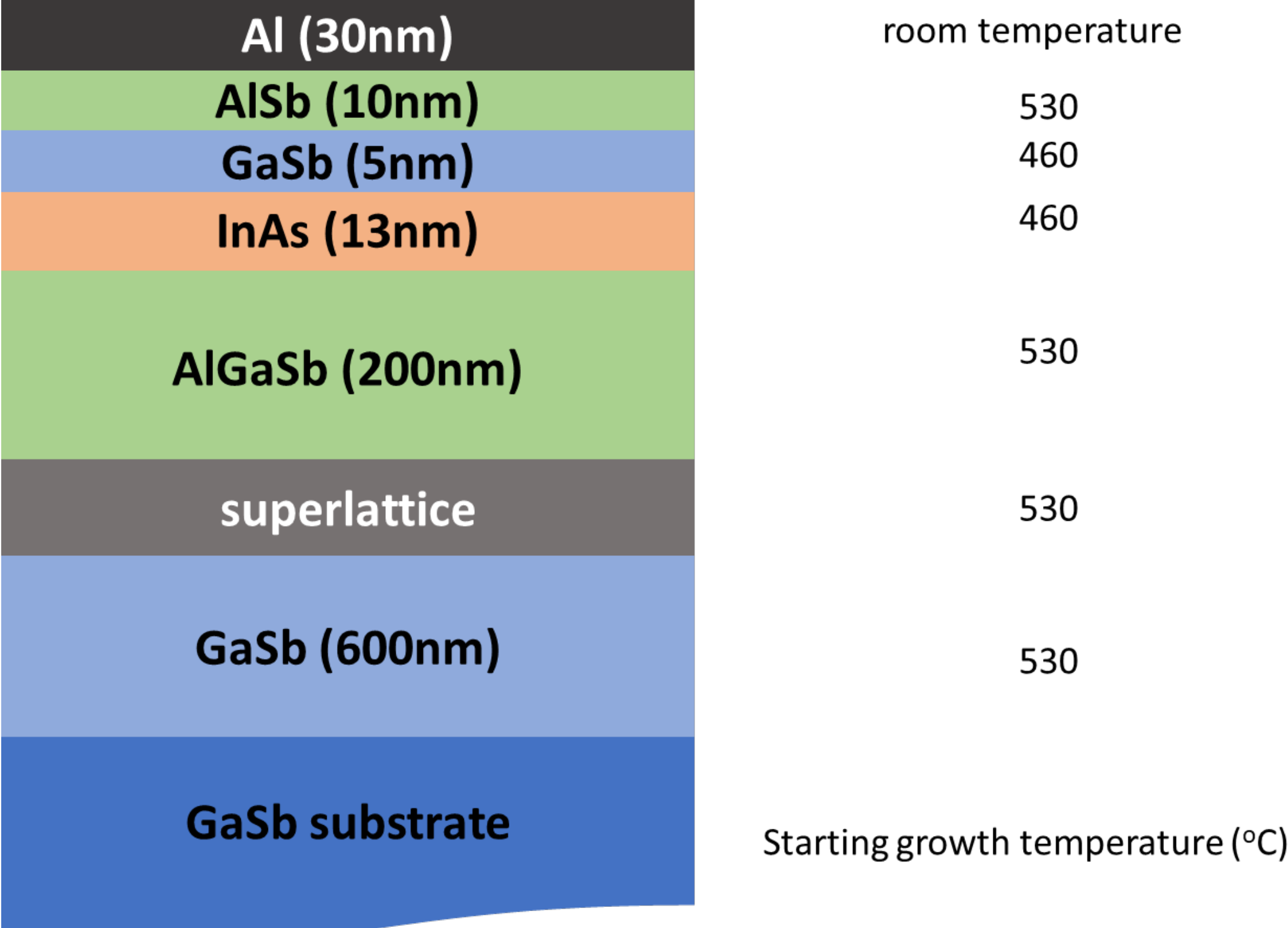

Fig. S1: growth structure and temperature for the first epi-Al/antimonide heterostructure.

Fig. S2 shows a schematic experimental flow of structure/device characterization for the measurements described in the main text. We first perform structural characterization of an as-grown wafer (in Fig. 1), then carry out a transport characterization of an epi-Al layer specimen to confirm that the epi-Al layer superconducts (in Fig. 2). Finally, we fabricate a Josephson junction from the epi-Al heterostructure to examine whether supercurrent states can be achieved in our Josephson junction (in Fig. 3).

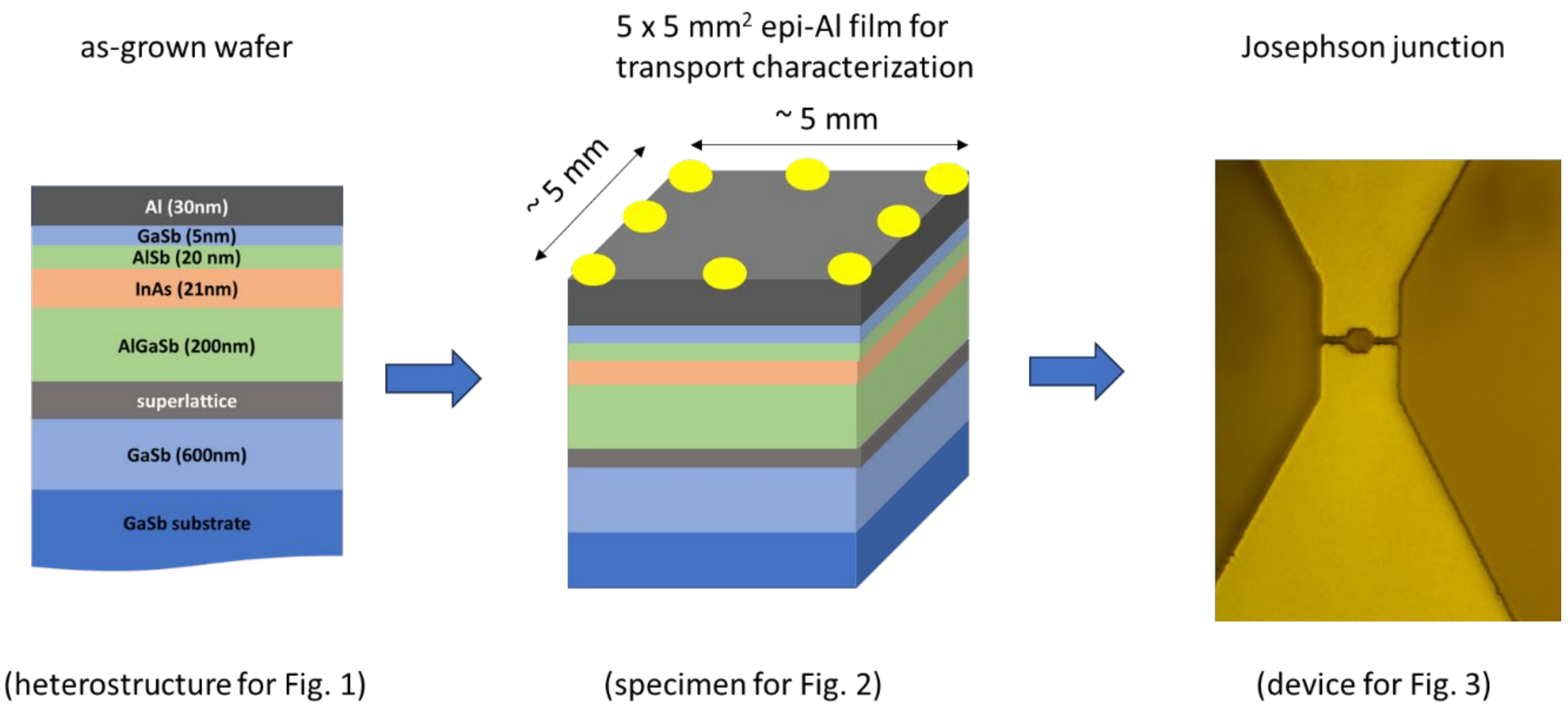

Figure S2: Experimental fabrication process for each measurement figure in the main text.

Figure S3 shows the setup for the V-I and dV/dI measurements. For the dV/dI measurements, a small ac current excitation $I_{ac}$ is added to the dc current $I_{dc}$.

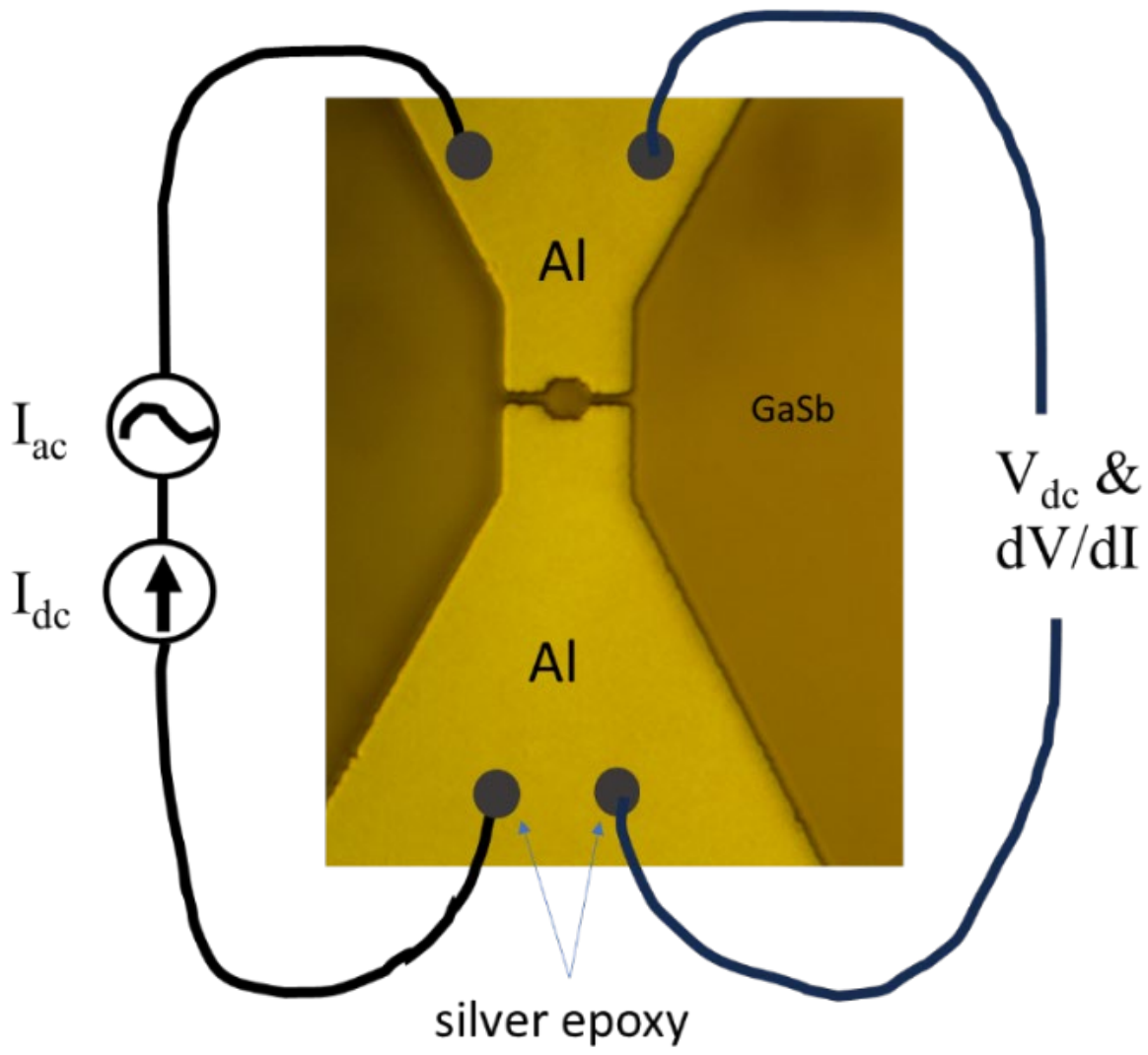

Figure S3: measurement setups for V-I and dV/dI.

To further characterize the superconducting properties of the epi-Al layer in the first heterostructure, we measure $R_{xx}$ as a function of magnetic field perpendicular to the epi-Al plane (Fig. S4a), from which the critical magnetic field $B_c$ is deduced. In Fig. S4b, we plot $B_c$ as a function of temperature. It follows the BCS formula $B_c = B_{c0}\times(1-(T/T_c)^2)$ [23], from which a zero-temperature critical magnetic field $B_{c0}$ = 43.4 mT and a critical temperature of 1.23 K (which is consistent with that in Fig. 2c) are obtained.

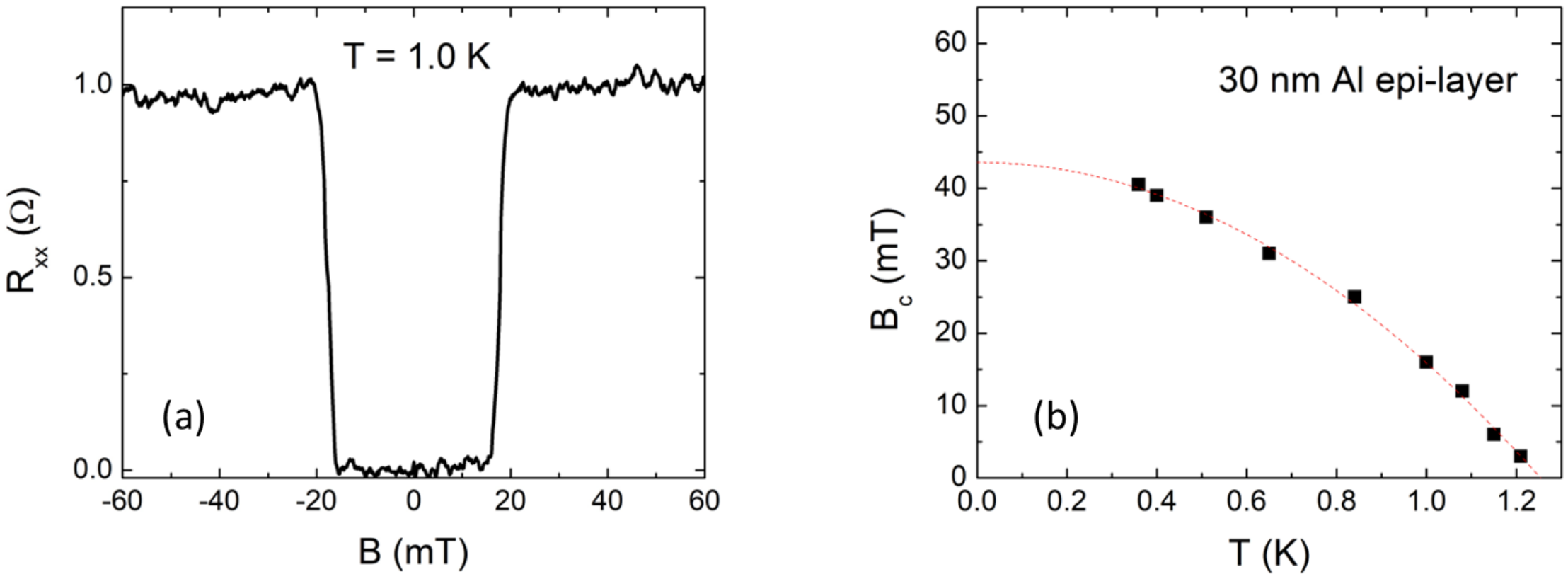

Figure S4: (a) Magnetoresistance of the epi-Al layer in the first heterostructure measured at T = 1.0 K. $R_{xx}$ = 0 W around B = 0 T. It jumps to a finite value at ~ ±16 mT. (b) Critical magnetic field ($B_c$) as a function of temperature in the epi-Al layer. The red line is a fit to the BCS formula.

Figure S5 shows the SEM images of epi-Al thin films for a typical wafer grown at room temperature and for the second heterostructure grown below 0 °C, respectively. It is clearly seen that the epi-Al layer in the wafer grown below 0 °C is much smoother.

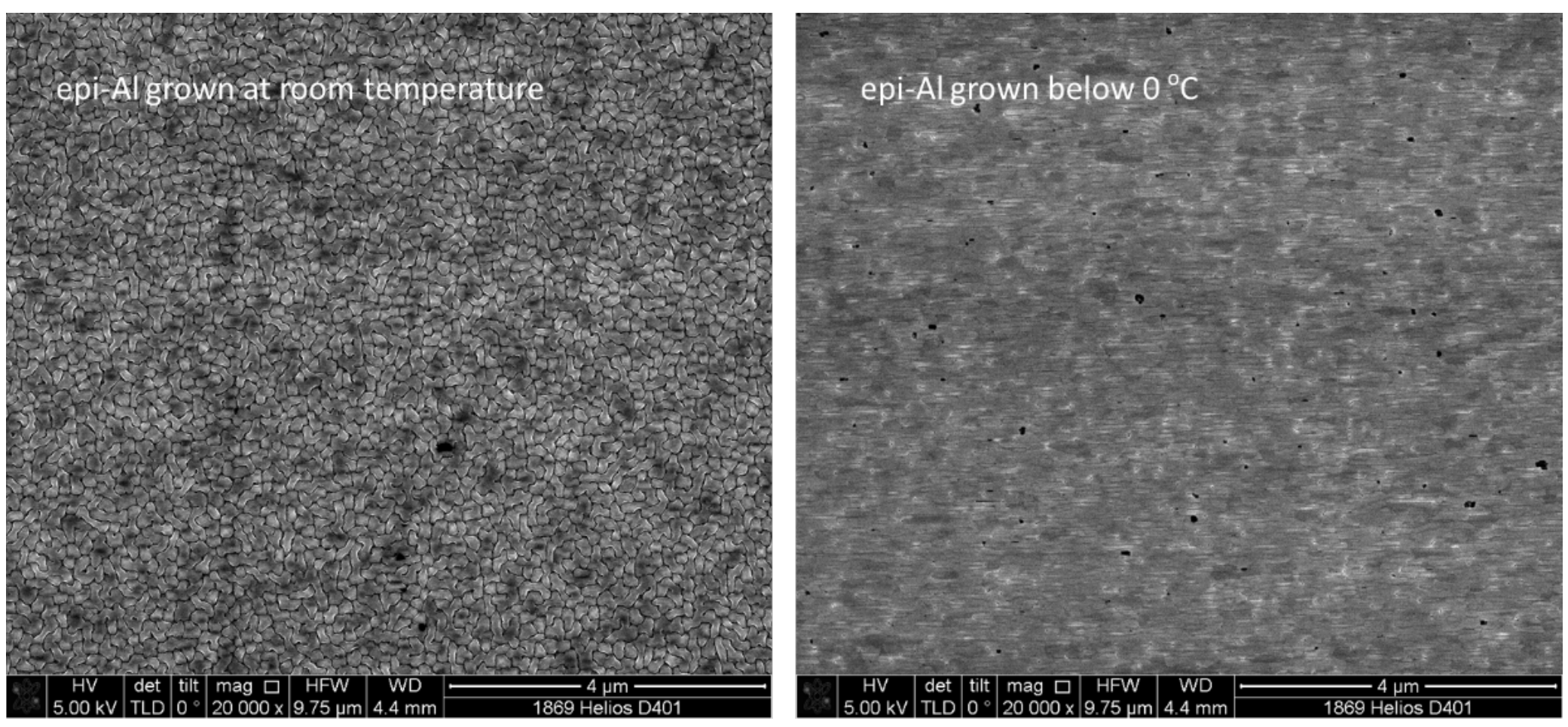

Figure S5: SEM images of epi-Al thin films for a typical wafer grown at room temperature and for the second heterostructure grown below 0 °C, respectively. The epi-Al layer in the wafer grown below 0 °C is much smoother.

Figure S6 shows $R_{xx}$ as a function of magnetic field perpendicular to the epi-Al plane in the second heterostructure from which a critical magnetic field $B_c \sim 5$ mT is obtained.

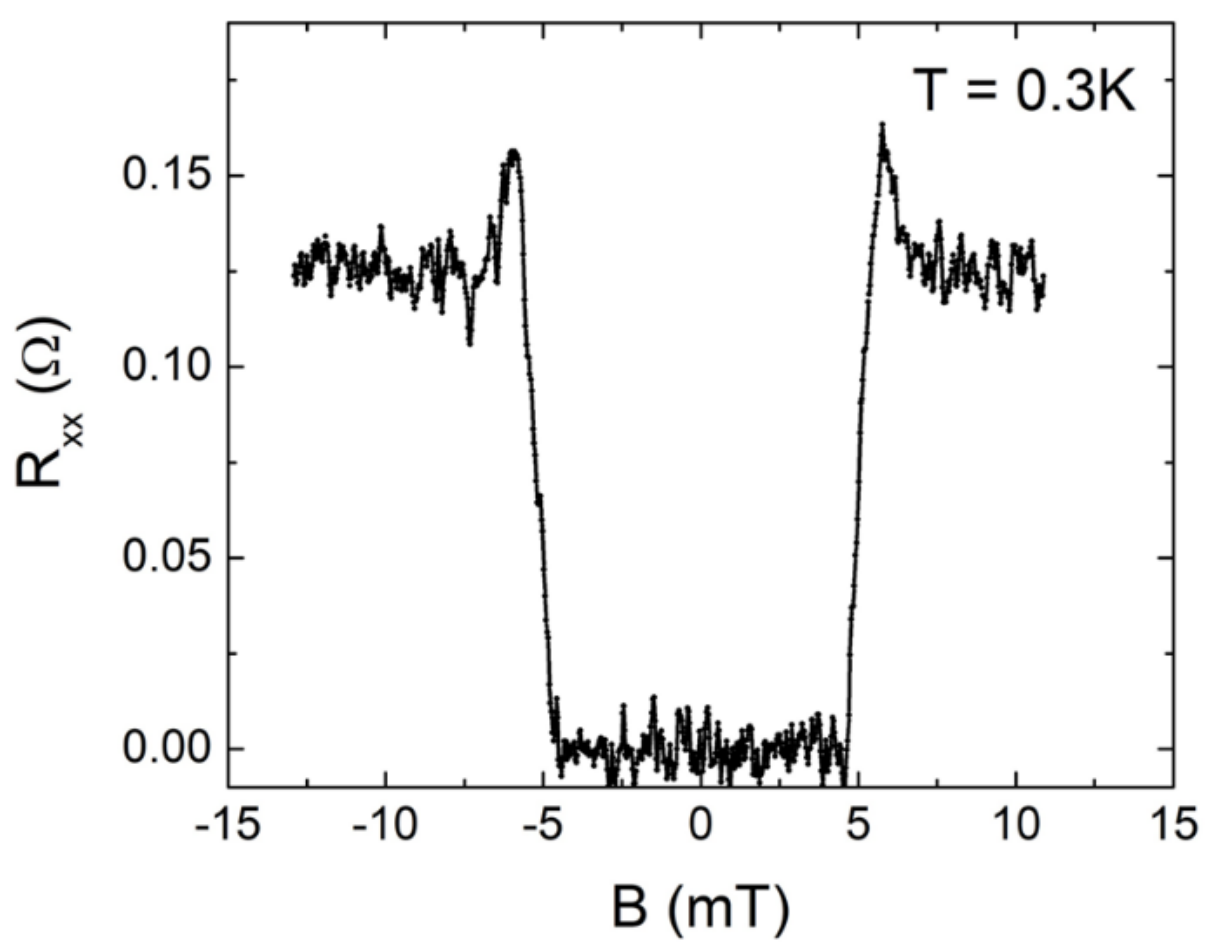

Figure S6: Magnetoresistance of the epi-Al layer in the second heterostructure measured at T = 0.3 K. $R_{xx} = 0\ \Omega$ around B = 0 T. It jumps to a finite value at $\sim \pm 5$ mT.

We also examine low temperature electronic characterizations of the 2DEG in the InAs QW without epi-Al. In Fig. S7b, we show the magnetoresistance $R_{xx}$ and Hall resistance $R_{xy}$ as a function of magnetic field; data are taken at the temperature of 0.3 K. Well-developed Shubnikov-de Haas oscillations and integer quantum Hall effect are observed in the low and high B regimes, respectively, demonstrating the high quality of the heterostructure.

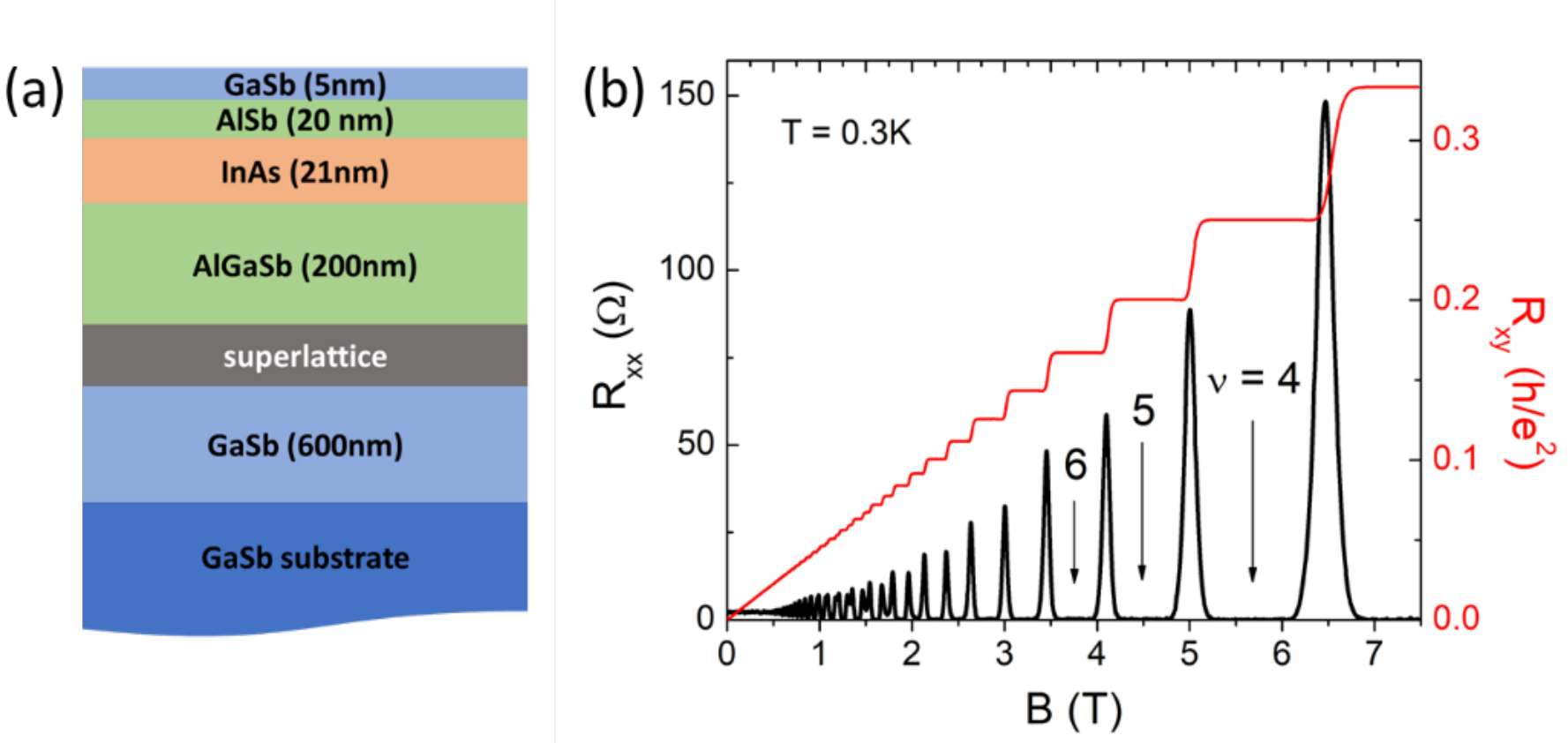

Fig. S7: (a) Growth structure of a high-mobility InAs quantum well structure. (b) Magnetoresistance $R_{xx}$ and Hall resistance $R_{xy}$ measured in this quantum well. Quantum Hall states at Landau level filling $\nu = 4, 5, 6$ are marked.

Figure S8 shows the V-I and dV/dI curves for the JJ in the main test at T ~ 0.9K in a different cool-down. Again, a multi-switching behavior is observed near the critical current (Fig. S8a). Due to this, dV/dI can assume negative values (Fig. S8b).

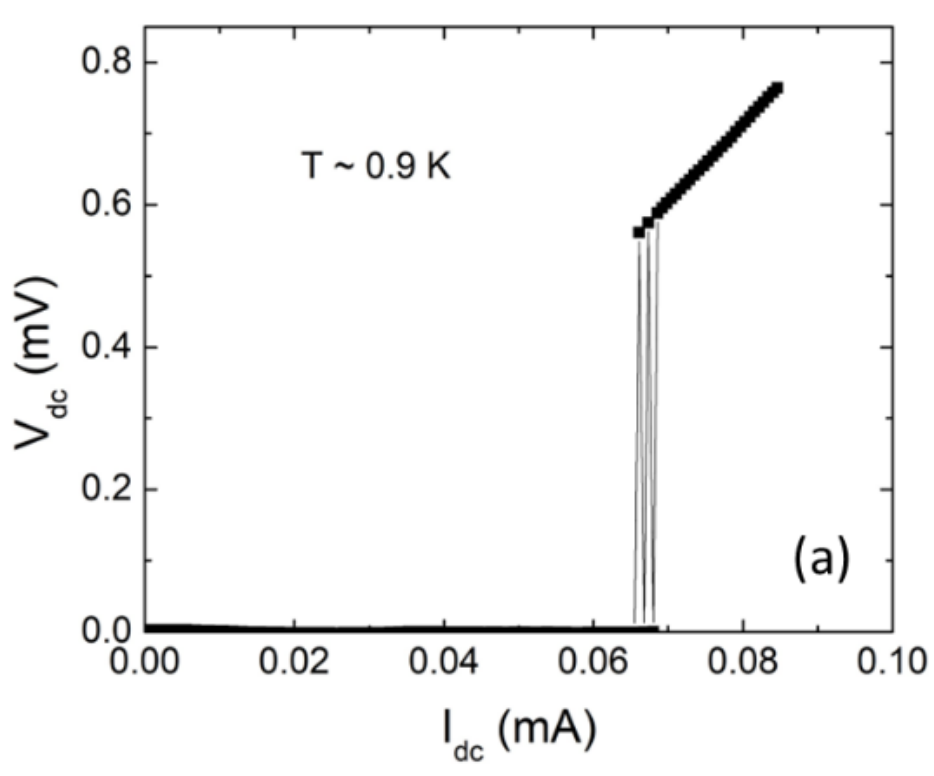

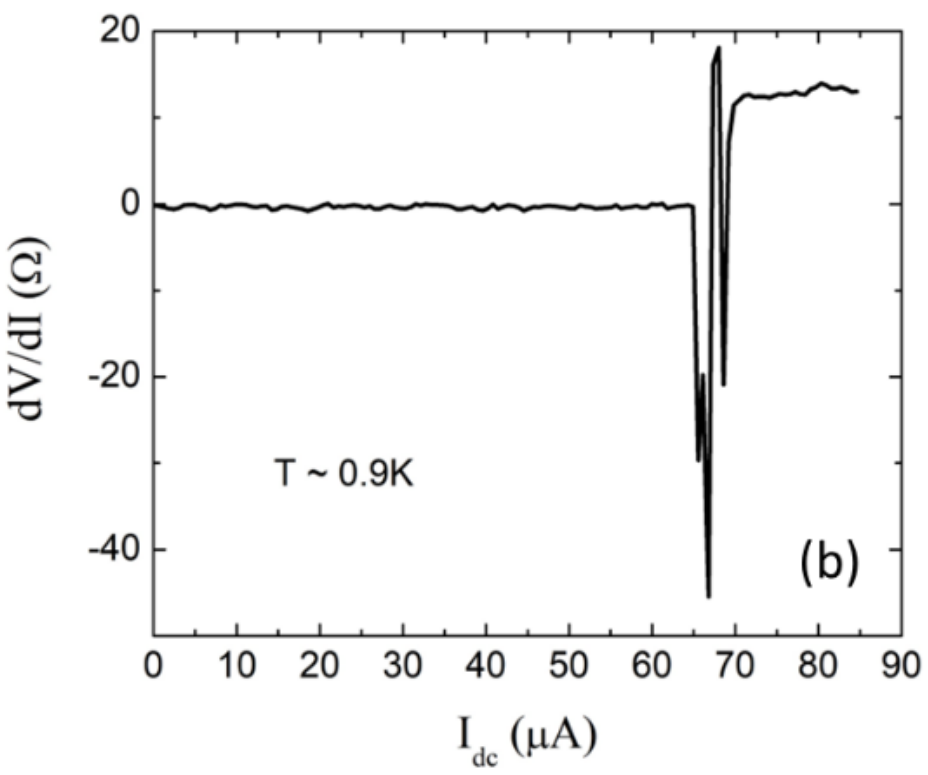

Fig. S8: (a) V-I curve. (b) dV/dI vs $I_{dc}$.